\documentclass[aps, prb,twocolumn,showpacs,superscriptaddress]{revtex4-1}

\usepackage{graphicx}% Include figure files
\usepackage{multirow}
\usepackage{ulem}
\usepackage[dvipdfm]{color}
\usepackage{amsmath,amssymb}

\def\mathi{\mathrm i}

\newcommand{\toffdiag}{\ensuremath{t_\mathrm{offdiag}}}
\newcommand{\tdiag}{\ensuremath{t_\mathrm{diag}}}
\newcommand{\Ja}{\ensuremath{J_\mathrm{a}}}
\newcommand{\Jb}{\ensuremath{J_\mathrm{b}}}
\newcommand{\Jc}{\ensuremath{J_\mathrm{c}}}
\newcommand{\JDM}{\ensuremath{J_\mathrm{DM}}}
\newcommand{\Ueff}{\ensuremath{U_\mathrm{eff}}}

\newcommand{\Dtri}{\ensuremath{\Delta_\mathrm{tri}}}

\newcommand{\awg}{\ensuremath{a_\mathrm{1g}}}

\newcommand{\egppm}{\ensuremath{e_\mathrm{g\pm}^\prime}}

\if0
\newcommand{\trsout}[1]{\textcolor{red}{\sout{#1}}}
\newcommand{\tbsout}[1]{\textcolor{blue}{\sout{#1}}}
\newcommand{\tgsout}[1]{\textcolor{green}{\sout{#1}}}
\newcommand{\tcsout}[1]{\textcolor{cyan}{\sout{#1}}}

\fi

\newcommand{\trsout}[1]{}
\newcommand{\tbsout}[1]{}
\newcommand{\tgsout}[1]{}
\newcommand{\tcsout}[1]{}

\begin{document}
\title{Spin-orbital frustration in pyrochlores $A_2$Mo$_2$O$_7$}

\author{Hiroshi Shinaoka}
\altaffiliation[Present address: ]{Theoretische Physik, ETH Z\"{u}rich, 8093 Z\"{u}rich, Switzerland}
\affiliation{Nanosystem Research Institute ``RICS'', National Institute of Advanced Industrial Science and Technology (AIST), Umezono, Tsukuba 305-8568, Japan}

\author{Yukitoshi Motome} 
\affiliation{Department of Applied Physics, University of Tokyo, Hongo, Bunkyo-ku, Tokyo 113-8656, Japan}

\author{Takashi Miyake}
\affiliation{Nanosystem Research Institute ``RICS'', National Institute of Advanced Industrial Science and Technology (AIST), Umezono, Tsukuba 305-8568, Japan}

\author{Shoji Ishibashi}
\affiliation{Nanosystem Research Institute ``RICS'', National Institute of Advanced Industrial Science and Technology (AIST), Umezono, Tsukuba 305-8568, Japan}

\date{\today}

%%%-----------------------------------------------------------------
\begin{abstract}
Electronic and magnetic properties of molybdenum pyrochlores $A_2$Mo$_2$O$_7$ are studied by the fully relativistic density-functional theory plus on-site repulsion ($U$) method, with focusing on the spin-glass insulating material Y$_2$Mo$_2$O$_7$.
We find that the system exhibits peculiar competition in energy between different magnetic states in the large-$U$ insulating region. 
The magnetic competition cannot be explained by the conventional picture based on the geometrical frustration of isotropic Heisenberg antiferromagnetic exchange interactions.
Through an analysis by using a generalized spin model, we find that the effective spin interactions are distinct from the simple Heisenberg form and strongly anisotropic in spin space.
We also reveal that they give rise to keen competition between antiferromagnetic and ferromagnetic states. 
The complex form of the magnetic interactions
indicates a crucial role of the orbital degree of freedom.
Analyzing a three-orbital Hubbard model, we clarify that the magnetic competition is tightly connected with orbital frustration in the $4d^2$ electronic configuration through the spin-orbital interplay.
The results challenge the conventional picture of the spin-glass behavior that attributes the origin to the geometrical frustration of purely antiferromagnetic exchange interactions.
\end{abstract}

% insert suggested PACS numbers in braces on next line

\maketitle

%%%-----------------------------------------------------------------
%%%  Introduction
%%%-----------------------------------------------------------------
\section{Introduction}
Spin and orbital degrees of freedom of electrons play a crucial role in strongly correlated electron systems.
The two degrees of freedom are coupled with each other via the strong Coulomb interaction and the relativistic spin-orbit interaction (SOI)~\cite{Kugel82}.
In general, the former interaction is important in 3$d$ transition metal compounds, while the latter is dominant in 5$d$ systems.
The spin-orbital interplay is a source of fascinating and intricate properties, 
such as complicated spin-orbital orderings~\cite{Tokura00} and topologically nontrivial states~\cite{Hasan10}.
Meanwhile, further intriguing situation is brought about by geometrical frustration of the lattice structure~\cite{Diep05,Lacroix11}.
Frustration suppresses a simple-minded ordering, and the residual spin and orbital fluctuations can induce interesting phenomena, such as heavy-fermion behavior and exotic orders. 

A family of pyrochlore oxides $A_2B_2$O$_7$ is a model system for studying the effects of spin-orbital interplay and geometrical frustration~\cite{Gardner10}.
In particular, compounds with $B$=Mn, Mo, Ir, and Os are interesting as they exhibit a metal-insulator transition (MIT) by changing temperature ($T$), pressure, and $A$-site cations. 
For instance, in 3$d$ systems with $B$=Mn, the importance of Coulomb interactions has been argued for the mechanism of MIT and giant magnetoresistance~\cite{Subramanian96,Shimakawa97}. 
On the other hand, for 5$d$ pyrochlores with $B$=Ir and Os, recent first-principles studies revealed 
that SOI plays a dominant role in determining their peculiar electronic and magnetic properties~\cite{Wan11,Shinaoka-CDO}.

\begin{figure}[!]
 \centering
 \includegraphics[width=.35\textwidth,clip]{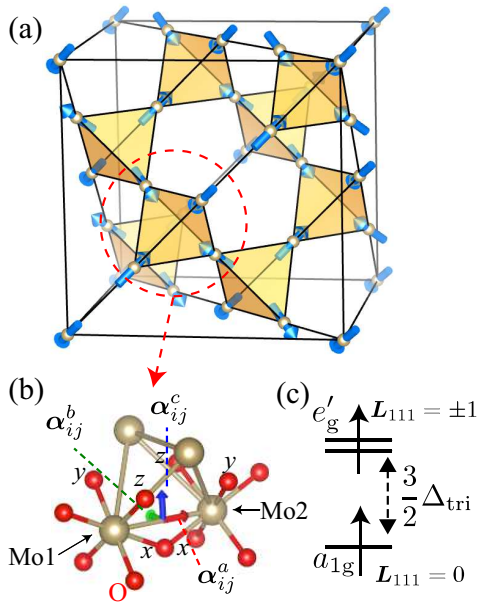}
 \caption{(color online). 
(a) Cubic unit cell of a pyrochlore lattice, composed of Mo atoms in $A_2$Mo$_2$O$_7$. 
The arrows represent the Mo spin moments in the 4in/4out magnetic structure, which are along the local [111] axes. 
(b) A primitive four-site unit cell, whose enlarged figure shows the local coordinates $xyz$ and the vectors $\boldsymbol{\alpha}_{ij}^k$ in the effective spin model in Eq.~(\ref{eq:spin-model}).
(c) The energy diagram shows the trigonal field splitting of $t_\mathrm{2g}$ level into $\awg$ and doubly degenerate $\egppm$ levels, each of which has a quantized angular momentum along the local [111] axis ($\boldsymbol{L}_{111}$).
}
 \label{fig:sys}
\end{figure}

Mo pyrochlores $A_2$Mo$_2$O$_7$ are of particularly interest as Mo 4$d$ electrons are subject to both strong Coulomb interactions and SOI. 
The system exhibits MIT by $A$-site substitution~\cite{Greedan87, Ali89, Katsufuji00} as well as external pressure~\cite{Iguchi09, Iguchi11}. 
The compounds with relatively large $A$-site ionic radii, e.g., Nd and Sm, show ferromagnetic (FM) metallic behavior at low $T$, while those with smaller ionic radii, such as Y, Dy, and Tb, are insulating and exhibit a spin-glass (SG) transition instead of conventional long-range ordering~\cite{Greedan86,Sato87,Raju92, Gingras96, Gingras97}. 
Electronic structure calculations showed that MIT is driven by the Coulomb interaction~\cite{Solovyev03}.
In addition, in the insulating phase, the Hund's coupling forms an effective $S=1$ spin, being coupled to the orbital moment in the trigonal crystal field of MoO$_6$ via SOI (see Fig.~\ref{fig:sys}). %[see Fig.~\ref{fig:system}].
However, the spin and orbital states were examined only for a few limited configurations, and the origin of SG was not specified.
On the other hand, the SG behavior has been studied by using spin-only models with isotropic Heisenberg exchange interactions~\cite{Saunders07,Andreanov10,Shinaoka10b,Shinaoka-LT2011}. 
There, the origin of SG was attributed to the antiferromagnetic (AFM) interactions under geometrical frustration.
Thus, it is still to be achieved a comprehensive understanding of the spin-orbital interplay and the peculiar magnetism.
Recently, neutron scattering experiments were performed for single crystals of an insulating compound Y$_2$Mo$_2$O$_7$~\cite{Silverstein13}.
They showed that diffuse magnetic scattering develops at low $T$ around [000] and ferromagnetic points such as [222],
indicating that the compound is not a simple isotropic Heisenberg antiferromagnet.
The importance of orbital degree of freedom and local lattice distortions was pointed out.
Thus, it is desired to carefully reexamine low-energy spin and orbital states by taking into account the Coulomb interaction and SOI on an equal footing.

In this paper, we investigate the electronic and magnetic properties in $A_2$Mo$_2$O$_7$ by the fully relativistic local spin density approximation(LSDA)+$U$ method.
The LSDA+$U$ result shows that, by increasing the electron correlation, the system exhibits MIT from a FM metal to an AFM insulator.
Remarkably, we find keen magnetic competition in the insulating phase,
which is not explained by the simple Heisenberg AFM model adopted in the previous studies.
Through the analysis by using a generalized spin model, we reveal that the competition originates from highly anisotropic effective spin interactions.
We also find that the system is in the competing regime between AFM and FM states. 
As a consequence, we show that the spin-spin correlation at finite temperature exhibits fluctuations of both AFM and FM components. 
To clarify the microscopic origin of the magnetic competition, we analyze the low-energy physics of a multi-orbital Hubbard model for the $t_{2g}$ orbitals. 
By introducing a control parameter in the realistic model obtained from the LSDA+$U$ analysis, 
we reveal that the magnetic competition is tightly connected with competing orbital states. 
Thus, our results renew the picture of the spin-glass insulating state in $A_2$Mo$_2$O$_7$; it is not a simple frustrated antiferromagnet but a spin-orbital frustrated Mott insulator.

This paper is organized as follows.
Section II is devoted to describing the method and system setup for the electronic structure calculations.
In Sec. III, we present results of the electronic structure calculations, and analyze the peculiar magnetic competition using the generalized spin model.
In Sec. IV, we discuss the microscopic origin of the magnetic competition through the analysis of the multi-orbital Hubbard model.
Summary is given in Sec. VIII.

%%%-----------------------------------------------------------------
%%%  Method
%%%-----------------------------------------------------------------
\section{Method}
We perform density-functional calculations with our computational code, QMAS (Quantum MAterials Simulator)~\cite{qmas}, using the projector augmented-wave method~\cite{Blochl94b} and the LSDA+$U$ method~\cite{Solovyev94,Dudarev98}.
The Perdew-Zunger formula~\cite{Ceperley80,Perdew81} is adopted for the LDA exchange-corrlation energy functional.
The relativistic effect including SOI is considered using two-component wave functions~\cite{Oda98,Kosugi_Au}.
The following calculations were done for a typical insulating material Y$_2$Mo$_2$O$_7$ using the experimental lattice structure:
the lattice constant $a=10.21$~\AA~\cite{Gardner99} and the so-called $u$ parameter $x(\mathrm{O}_1)=0.33821$~\cite{Reimers1988390}.
Every MoO$_6$ tetrahedron is compressed along the local [111] axis (trigonal distortion) for $x(\mathrm{O}_1)>0.3125$ [see Fig.~\ref{fig:sys}(b)].
We adopt a primitive unit cell with four Mo atoms, as shown in Fig.~\ref{fig:sys}(a). 
Experimentally, the magnetic and electronic properties vary systematically with the $A$-site ionic radius~\cite{Greedan87, Ali89, Katsufuji00}, which is regarded as the bandwidth control, namely, the control of electron correlation. 
We discuss such a systematic change by controlling the strength of electron correlation, $\Ueff~(\equiv U-J)$~\cite{Solovyev94,Dudarev98}.
Brillouin-zone integrations were performed using the improved tetrahedron method~\cite{Blochl94}.
We confirmed that results are converged with respect to planewave cutoff energy and the number of $k$ points.
In the following calculations, we use a planewave cutoff energy of 40~Ry.
The $k$ mesh is typically $8\times 8\times 8$ and $4\times 4\times 4$ for metallic and insulating phases, respectively.

%%%-----------------------------------------------------------------
%%%  LSDA+U results
%%%-----------------------------------------------------------------
\begin{figure*}[!]
 \centering
 \includegraphics[width=.75\textwidth,clip]{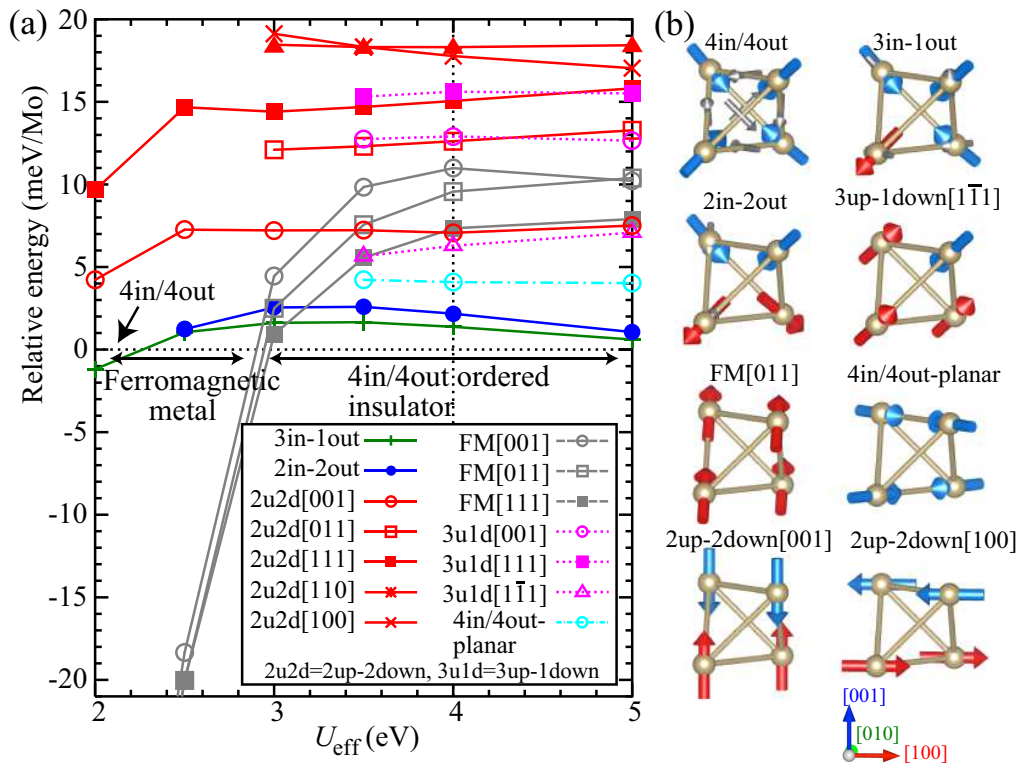}
 \caption{(color online). 
(a) LSDA+$U$ results for $U_\mathrm{eff}$ dependences of energies for various $\boldsymbol{q}=0$ magnetic states.
The magnetic patterns are shown in (b).
Energy is measured from that for the 4in/4out state.
%The arrows for the data at $\Ueff =4$ eV denote the energies calculated by the model (\ref{eq:spin-model}). 
In the 4in/4out state in (b), small gray arrows denote the DM vectors $\boldsymbol{\alpha}^\mathrm{DM}_{ij}$ in the model (\ref{eq:spin-model}).
}
 \label{fig:pd}
\end{figure*}
\section{Magnetic competition in the insulating phase}
\subsection{Electronic structure calculation}
Figure~\ref{fig:pd}(a) shows the energies of various types of magnetic structures as functions of the Coulomb repulsion $\Ueff$.
Here we performed self-consistent calculations with constraints on the directions of Mo spin moments depending on each magnetic structure shown in Fig.~\ref{fig:pd}(b).
The system exhibits MIT at $\Ueff \simeq 3$~eV from a FM metal to a magnetic insulator while increasing $\Ueff$, being consistent with the trend in the $A$-site substitution in $A_2$Mo$_2$O$_7$.

In the insulating region for $\Ueff\gtrsim3$~eV, the 4in/4out order is the most stable among the magnetic structures considered [see Figs.~\ref{fig:sys}(a) and \ref{fig:pd}(b)]; the Mo spin moment is 1.54 $\mu_\mathrm{B}$ at $\Ueff=4$ eV~\footnote{The reduction of the spin moment from the ideal value may be due to the proximity effect of MIT and the hybridization with O $p$ orbitals}, indicating the formation of an effective $S=1$ spin by the Hund's coupling under the trigonal crystal field.
As illustrated in Fig.~\ref{fig:sys}(c), $t_\mathrm{2g}$ level splits into $\awg$ and doubly degenerate $\egppm$ levels, being half/quarter filled respectively in the 4$d^2$ configuration~\cite{Solovyev94}.
There are, however, many other low-energy insulating states, with their relative energies being almost independent of $\Ueff$.
In particular, the 3in--1out and 2in--2out states are energetically very close to the 4in/4out ground state.
This competition is suggestive of the SG behavior because the energy difference is comparable to or even smaller than the energy scale of the SG transition temperature 20--25 K~\cite{Greedan86,Sato87,Raju92, Gingras96, Gingras97}.
The fact that these three spin states have a low energy is consistent with a substantial local [111] easy-axis spin anisotropy, which was found in the previous study~\cite{Solovyev03}.
Nevertheless, a simple AFM Heisenberg model with the easy-axis anisotropy adopted in Ref.~\onlinecite{Solovyev03} cannot account for the energy spectrum in Fig.~\ref{fig:pd}(a).
For example, FM[111]([011]) is always substantially lower in energy than 2up--2down[111]([011]), contrary to the expectation for the AFM model.
This strongly suggests the existence of further intricate spin interactions.

\subsection{Analysis by a generalized spin model}
To clarify the origin of this peculiar magnetic competition, we consider a generalized spin model including all the symmetry-allowed pairwise interactions between nearest-neighbor (NN) spins in addition to the single-ion anisotropy~\cite{Jackeli09,Kim09}.
For simplicity, we restrict the consideration to classical spins.
The Hamiltonian is written in the form
\begin{eqnarray}
	&&\mathcal{H}=\sum_{\langle ij\rangle } \sum_{k=a,b,c} J_k (\boldsymbol{S}_i \cdot \boldsymbol{\alpha}_{ij}^k) (\boldsymbol{S}_j \cdot \boldsymbol{\alpha}_{ij}^k)\nonumber \\
	&&+J_\mathrm{DM}\sum_{\langle ij\rangle, i<j } \boldsymbol{\alpha}^\mathrm{DM}_{ij} \cdot (\boldsymbol{S}_i \times \boldsymbol{S}_j)-D\sum_i  \left(\boldsymbol{S}_i \cdot \boldsymbol{\alpha}^{111}_i\right)^2,\label{eq:spin-model}
\end{eqnarray}
where $J_k$ ($k=a, b, c$) are anisotropic NN exchange couplings between Mo spins $\boldsymbol{S}_i$ and $\boldsymbol{S}_j$;
$\boldsymbol{\alpha}_{ij}^k$ are normalized vectors along cubic axes on NN bonds [see Fig.~\ref{fig:sys}(b)]. 
We take $|\boldsymbol{S}_i|=1$.
The second term denotes the Dzyaloshinsky-Moriya (DM) interaction~\cite{Elhajal05}.
The DM vectors $\boldsymbol{\alpha}^\mathrm{DM}_{ij}$ are shown in Fig.~\ref{fig:pd}(b) ($|\boldsymbol{\alpha}^\mathrm{DM}_{ij}|=1$).
The third term represents the single-ion anisotropy $D$; $\boldsymbol{\alpha}^{111}_i$ is a normalized vector along the local [111] axis.

We determine the parameters in the model (\ref{eq:spin-model}) by fitting the LSDA+$U$ relative energies for all the magnetic states in Fig.~\ref{fig:pd}(a).
We find that all the levels are well explained by a highly anisotropic spin model with $\Ja>0$, $\Jb<0$, $\Jc<0$, $\JDM<0$, and $D>0$. 
For instance, at $\Ueff$=4 eV, we obtained the following estimates:
\begin{eqnarray}
\left\{
\begin{array}{l}
\Ja = 3.97 \pm 0.22{\rm ~meV} \\
\Jb = -3.00 \pm 0.23{\rm ~meV} \\
\Jc = -4.91 \pm 0.14{\rm ~meV} \\
\JDM = -4.00 \pm 0.18{\rm ~meV} \\
D = 17.52 \pm 0.78{\rm ~meV}.
\end{array}
\right.
\label{eq:Jparameters}
\end{eqnarray}
A comparison of energies between the LSDA+$U$ results and those from the model (\ref{eq:spin-model}) with the parameters in Eqs.~(\ref{eq:Jparameters}) is shown in Table~\ref{tbl:param}.
The LSDA+$U$ energies are reproduced within errors of 2 meV/Mo for all the magnetic states.

The resultant effective spin model explains the magnetic competition of low-energy states in Fig.~\ref{fig:pd}(a);
in the presence of the substantial $D>0$, the AFM $\Ja$ favors 4in/4out rather than 2in--2out and 3in--1out, while the FM $\Jb$, $\Jc$ and the negative $\JDM$ do the opposite.
The spin model is in keen competition between AFM and FM:
In fact, the ground state of the spin model sensitively changes from a 4in/4out AFM state to a 2in--2out-like FM state by a few \% modification of the model parameters, as demonstrated in the next subsection.
\begin{table}[h]
	\centering
	\begin{tabular}{l|ccc}
		\hline\hline
		\multirow{2}{*}{Configuration} & \multicolumn{3}{c}{Energy (meV/Mo)} \\
		& $E_{\mathrm{LSDA}+U}$ & $E_\mathrm{model}$ & $\Delta E$ %$E_\mathrm{model}-E_{\mathrm{LDA}+U}$ 
		\\
		\hline\hline
                4in/4out & 0.00 & 0.00 &  0.00\\
                3in--1out & 1.38 & 1.54 &  0.16\\
                2in--2out & 2.17 & 2.05 &  -0.12\\
                4in/4out-planar & 4.08 & 3.67 &  -0.42\\
                3up--1down[1$\bar{1}$1] & 6.27 & 6.25 &  -0.02\\
                2up--2down[001] & 7.07 & 7.34 &  0.27\\
                FM[111] & 7.34 & 9.28 &  1.94\\
                FM[011] & 9.57 & 9.28 &  -0.29\\
                FM[001] & 10.99 & 9.28 &  -1.71\\
                2up--2down[011] & 12.62 & 12.73 &  0.12\\
                3up--1down[001] & 12.90 & 13.22 &  0.32\\
                2up--2down[111] & 15.06 & 14.53 &  -0.52\\
                3up--1down[111] & 15.62 & 15.54 &  -0.07\\
                2up--2down[100] & 17.77 & 18.13 &  0.36\\
                2up--2down[110] & 18.31 & 18.13 &  -0.18\\
		\hline\hline
	\end{tabular}
	\caption{Energy comparison for various magnetic configurations for $\Ueff=4$ eV.
	$E_{\mathrm{LSDA}+U}$ denotes the results obtained by the LSDA+$U$ calculation at $\Ueff=4$~eV in Fig.~\ref{fig:pd}(a).
	$E_\mathrm{model}$ denotes the calculated energies by the generalized model in Eq.~(\ref{eq:spin-model}) with the parameters in Eqs.~(\ref{eq:Jparameters}).
	}
	\label{tbl:param}
\end{table}

\subsection{Competition between antiferromagnetic and ferromagnetic states: spin-spin correlation}\label{sec:ss}
For the isotropic AFM spin model employed in the previous studies, 
the macroscopic ground-state degeneracy leads to peculiar spin correlations at finite $T$. 
For example, pinch points are seen in the spin structure factor at the $\Gamma$ point and equivalent points~\cite{Zinkin95,Zinkin97}.
However, the highly anisotropic nature of the effective spin model (\ref{eq:spin-model}) indicates that the model has qualitatively different aspects in its finite-$T$ spin correlations.
In this subsection, we calculate finite-$T$ spin correlations of the model (\ref{eq:spin-model}) by classical Monte Carlo (MC) simulation, and discuss effects of the competing anisotropic AFM and FM interactions.

In the following MC simulations, we calculate the spin structure factor defined as
\begin{eqnarray}
  S(\boldsymbol{q}) = \frac{1}{N_\mathrm{s}}\sum_{ij} \langle \boldsymbol{S}_i \cdot \boldsymbol{S}_j \rangle e^{\mathrm{i} \boldsymbol{q} \cdot (\boldsymbol{R}_i - \boldsymbol{R}_j)},
\end{eqnarray}
where $N_{\mathrm s}$ is the total number of spins, $\boldsymbol{R}_i$ is the position of $\boldsymbol{S}_i$ and $\langle \cdots \rangle$ is a thermal average.
We use systems of $8\times 8\times 8$ in terms of the cubic unit cell, i.e., $N_{\mathrm s} = 16\times 8^3 =8192$.
The number of MC steps is $10^5$.
The data are averaged over 16 independent MC runs starting with different random seeds.

\begin{figure}[!]
 \centering
 \includegraphics[width=.35\textwidth,clip]{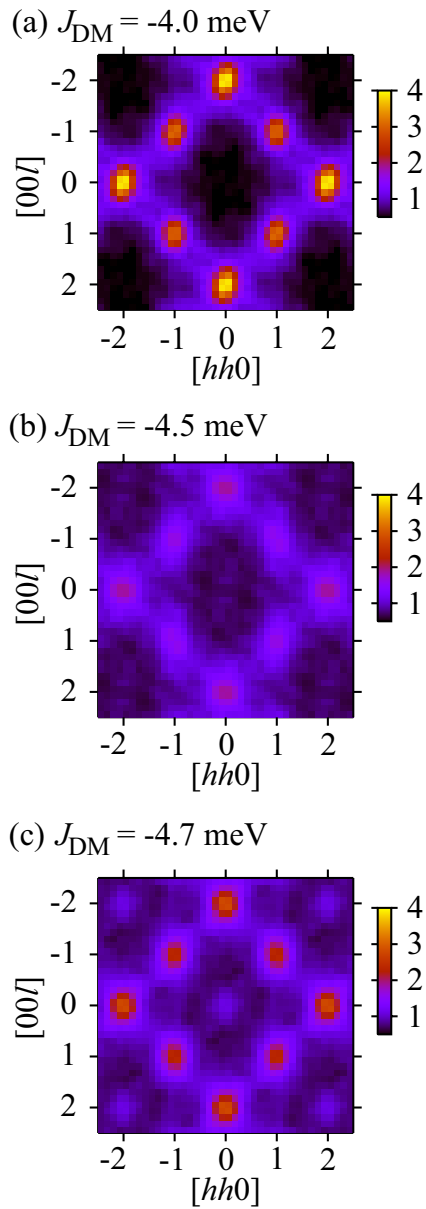}
 \caption{(color online). 
(a) The spin structure factor $S(\boldsymbol{q})$ for the model (\ref{eq:spin-model}) with the parameters in Eqs.~(\ref{eq:Jparameters}).
(b) and (c) show the results when varying the values of $\JDM$ to $-4.5$~meV and $-4.7$~meV, respectively.
The ground state changes from the 4in/4out AFM state to a 2in--2out-like FM state at $\JDM\simeq -4.5$ meV as decreasing $\JDM$.
The results are calculated in the finite-$T$ PM phases slightly above the magnetic transition temperature: (a) 22.5~K, and (b) and (c) 15~K.
 }
 \label{fig:spin-spin-correlation}
\end{figure}

Figure~\ref{fig:spin-spin-correlation}(a) shows $S(\boldsymbol{q})$ calculated at finite $T$ in the PM phase for the anisotropic spin model (\ref{eq:spin-model}) with the parameters in Eqs.~(\ref{eq:Jparameters}).
Since the ground state is the 4in/4out AFM state for these parameters, 
$S(\boldsymbol{q})$ shows fluctuations at the same positions as the Bragg peaks in the ground state, e.g., at [220], [111], [022].
This is clearly distinguished from the pinch-point structure for the NN antiferromagnet.
As discussed above, the ground state of the model sensitively turns into the 2in--2out-like FM state by a small change in the model parameters.
To investigate effects of the characteristic spin-spin correlation brought by the AFM-FM competition,
here, we vary the value of $\JDM$ systematically to to change the ground state.
We find that the system is in the 4in/4out state for $\JDM\lesssim -4.5$~meV, while it becomes the 2in-2out-like state for $\JDM\gtrsim -4.5$~meV. 
As shown in Figs.~\ref{fig:spin-spin-correlation}(b) and \ref{fig:spin-spin-correlation}(c), when the ground state turns into the 2in--2out-like FM state, broad spots appear at [222] and equivalent points at finite $T$, corresponding to FM fluctuations in a tetrahedron~\footnote{$e^{\mathrm{i}\boldsymbol{q}R}$ with $\boldsymbol{q}$=[222] changes the phase by $2\pi$ between neighboring Kagome and triangular planes in the pyrochlore lattice.}.
These competing AFM and FM spin fluctuations are characteristic to the model (\ref{eq:spin-model}).
Our results are suggestive of the recent observation of diffuse magnetic scattering at FM points such as [000] and [222] in Y$_2$Mo$_2$O$_7$~\cite{Silverstein13}.

\section{Microscopic origin of magnetic degeneracy}
The surprisingly anisotropic nature of the effective spin model in Eq.~(\ref{eq:spin-model}) clearly indicates that the orbital degree of freedom plays a substantial role.
In particular, the magnetism may depend on the orbital state of doubly-degenerate $\egppm$ levels via SOI because $\egppm$ have orbital moments along the local [111] axes [see Fig.~\ref{fig:sys}(c)].
In this section, we clarify the microscopic origin of the magnetic competition by analyzing the relation between the magnetism and orbital state.

\subsection{Analysis of a multi-orbital Hubbard model}
\begin{figure}[h]
 \centering
 \includegraphics[width=.5\textwidth,clip]{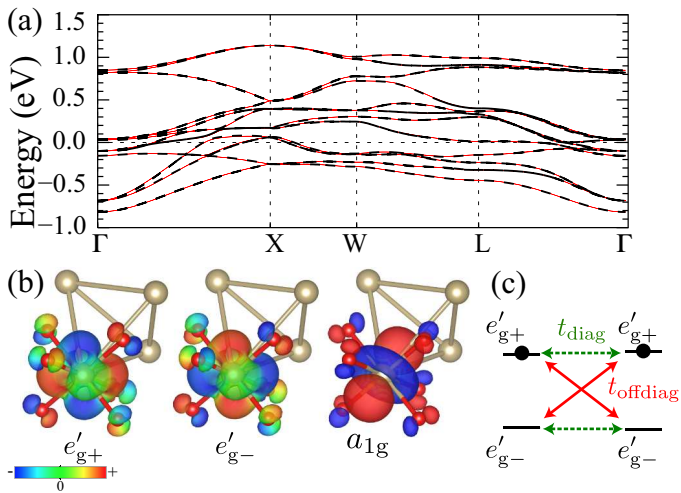}
 \caption{(color online). 
 (a) Nonmagnetic band structure (solid curves) calculated for $\Ueff$=0.
 The broken curves represent the fit by MLWF. The energy is measured from the Fermi level.
 (b) Density distribution of the MLWF localized at a Mo site obtained for $\Ueff$=0.
 We take the [001] axis as the quantization axis.
 The surface coloring represents the imaginary part ($\egppm$) and real part ($\awg$) of the wavefunctions of the majority spin component, respectively.
 Note that the $\egppm$ MLWF are complex conjugates of each other.
 (c) The orbital off-diagonal ($\toffdiag$) and diagonal ($\tdiag$) transfers between $\egppm$ orbitals on NN Mo atoms.
 }
 \label{fig:wannier}
\end{figure}

To clarify the role of orbitals, we analyze a three-orbital Hubbard model for the $t_{2g}$ orbitals.
The Hamiltonian considered is 
\begin{eqnarray}
\!\!\!\!\!&&\mathcal{H}=\sum_{\langle ij\rangle}\sum_{\alpha\beta}(t_{ij\alpha\beta} c_{i\alpha\sigma}^\dagger c_{j\beta\sigma^\prime} + \mathrm{H.c.})+\sum_i \Big(\zeta\hat{l_i}\cdot \hat{s_i}+\nonumber \\
\!\!\!\!\!&&\frac{2\Dtri}{3}\!\!\!\!\sum_{\alpha=e_\mathrm{g\pm}^\prime}\!\!c_{i\alpha}^\dagger c_{i\alpha}+\!\!\!\!\!\!\!\!\sum_{\alpha\beta\alpha^\prime\beta^\prime\sigma\sigma^\prime}\!\!\!\!\!\frac{U_{\alpha\beta\alpha^\prime\beta^\prime}}{2}\tilde{c}^\dagger_{i\alpha\sigma}\tilde{c}^\dagger_{i\beta\sigma^\prime}\tilde{c}_{i\beta^\prime\sigma^\prime}\tilde{c}_{i\alpha^\prime\sigma}\Big),\label{eq:Hubbard}
\end{eqnarray}
where $i$, $j$ are indices of Mo sites, and $\alpha$, $\beta$ those of orbitals.
Spins $\sigma,\sigma^\prime$ are quantized in the [001] axis.
$c_{i\alpha}$ and $c^\dagger_{i\alpha}$ are the annihilation and creation operators for $\awg$ and $\egppm$ orbitals, while $\tilde{c}_{i\alpha}$ and $\tilde{c}^\dagger_{i\alpha}$ those for $d_{xy}$, $d_{yz}$, and $d_{zx}$ orbitals.
Here, we define the $\awg$ and $\egppm$ bases as
\begin{eqnarray}
  |\awg\rangle &=& \frac{1}{\sqrt{3}}(1,1,1),\\
  |\egppm\rangle &=& \frac{1}{\sqrt{3}}(e^{\pm 2\pi\mathi/3},1,e^{\mp 2\pi\mathi/3}),
\end{eqnarray}
respectively, in the basis of $d_{xy}$, $d_{yz}$, and $d_{zx}$ orbitals.
The first term in Eq.~(\ref{eq:Hubbard}) denotes transfers between NN Mo atoms.
The second term describes the LS coupling $\zeta$, the trigonal distortion $\Dtri$, and the rotationally symmetric on-site Coulomb interactions $U_{\alpha\beta\alpha^\prime\beta^\prime}$, respectively.
The Coulomb interaction is parameterized as $U_{\alpha\alpha\alpha\alpha}=U$, $U_{\alpha\beta\alpha\beta}=U-2J_\mathrm{H}$, $U_{\alpha\beta\beta\alpha}=U_{\alpha\alpha\beta\beta}=J_\mathrm{H}$ ($\alpha\neq \beta$), where $U$ is the on-site repulsion and $J_\mathrm{H}$ the Hund's coupling, respectively.
We use the realistic values of $\zeta$ and $\Dtri$ obtained by the maximally localized Wannier function (MLWF) analysis~\cite{Marzari97,Souza01} at $\Ueff=0$ in a manner similar to that in Ref.~\onlinecite{Solovyev06}: $\Dtri=0.25$ eV and $\zeta=0.085$ eV.
Figures~\ref{fig:wannier}(a) and \ref{fig:wannier}(b) show the band structure and the MLWF, respectively.
As seen in Fig.~\ref{fig:wannier}(a), the MLWF well reproduces the band structure of the $t_\mathrm{2g}$ manifold.
%In Fig.~\ref{fig:wannier}(a), we see that the MLWF well reproduces the band structure of the $t_\mathrm{2g}$ manifold.
%The MLWF's are shown in Fig.~\ref{fig:wannier}(b).
\begin{figure}[h]
 \centering
 \includegraphics[width=.5\textwidth,clip]{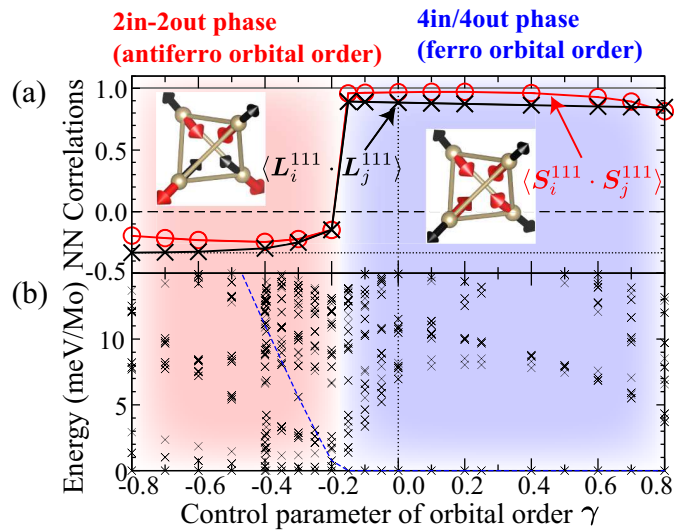}
 \caption{(color online). 
 The phase diagram of the three-orbital Hubbard model given in Eq.~(\ref{eq:Hubbard}).
 (a), (b) The NN correlation functions and the excitation spectrum.
 The parameter $\gamma$ controls the ratio $|\toffdiag/\tdiag|$ (see the text).
 The solid, broken, and dotted horizontal lines in (a) denote the NN correlations of $\boldsymbol{L}^{111}$ for the 4in/4out, the 3in--1out, and the 2in--2out state in the classical limit, respectively. 
 The insets in (a) schematically show the 4in/4out state (right) and the 2in--2out state (left). 
 Spin and orbital moments are denoted by the thick (red) and thin (black) arrows, respectively.
The crosses connected by the broken line in (b) denote the 4in/4out energy.
 }
 \label{fig:hubbard}
\end{figure}

To clarify the role of orbitals,
we introduce a control parameter in the transfer integrals $t_{ij\alpha\beta}$ obtained by the MLWF analysis as follows.
The ordering of $\egppm$ orbitals and the magnetism strongly depend on the relative magnitude of the orbital diagonal and off-diagonal NN transfers~\cite{Motome11}, denoted by $\tdiag$ and $\toffdiag$, respectively [see Fig.~\ref{fig:wannier}(c)]
That is, under strong Coulomb interactions, $\tdiag$ ($\toffdiag$) favors an antiferro (ferro) orbital alignment to gain the second-order perturbation energy.
To see how the orbital ordering affects the magnetism,
we control the ratio $|\toffdiag/\tdiag|$ by taking 
\begin{eqnarray}
  \toffdiag\rightarrow (1+\gamma) \toffdiag, \quad \tdiag\rightarrow (1-\gamma) \tdiag
\end{eqnarray}
with a control parameter $\gamma$.
The case with $\gamma$=0 corresponds to the MLWF estimate: $|\toffdiag|=0.145$ eV and $|\tdiag|=0.102$ eV.

For simplicity, we calculate the ground state and excited spectrum of the model~(\ref{eq:Hubbard}) as follows.
First, we calculate the multiplet structure for a Mo atom for $\zeta=0$, that is, all eigenenergies and eigenstates.
Then, using the multiplet basis, we construct a perturbative Hamiltonian in terms of $t_{ij}$ and $\zeta$ up to the lowest order, i.e., $O(t_{ij}^2)$ and $O(\zeta)$ for a primitive unit cell under the periodic boundary condition.
Diagonalizing the perturbative Hamiltonian, we obtain the ground state and excited states.
The results shown below do not change qualitatively for the choice of $U$ and $J_\mathrm{H}$ as long as $J_\mathrm{H}>\Dtri$; the condition is assured by the fact that the ground state is not spin-singlet.

Figure~\ref{fig:hubbard}(a) shows the results for NN spin and orbital correlations in the ground state at $U=4$ eV and $J_\mathrm{H}=0.5$ eV.
The correlation functions are calculated by taking the average over NN bonds, 
and $\boldsymbol{L}^{111}$ and $\boldsymbol{S}^{111}$ are the orbital and spin moments projected on the local [111] axes, respectively; 
namely, e.g., $\langle\boldsymbol{S}_i^{111}\cdot\boldsymbol{S}_j^{111}\rangle$ becomes positive for the AF 4in/4out state.
As shown in Fig.~\ref{fig:hubbard}(a), for $\gamma>\gamma_\mathrm{c}\simeq -0.17$, the NN correlations of $\boldsymbol{L}^{111}$ and $\boldsymbol{S}^{111}$ are positive and the values are close to those of the 4in/4out state. 
With decreasing $\gamma$, the two NN correlations change the signs simultaneously at $\gamma=\gamma_\mathrm{c}$.
For $\gamma<\gamma_\mathrm{c}$, their values are close to those of the 2in--2out state.

The change of the spin and orbital states by $\gamma$ is understood as follows.
First, the orbital moment $\boldsymbol{L}$ is polarized along the local [111] axis at each site under the trigonal crystal field [see Fig.~\ref{fig:sys}(c)].
Then, the polarized $\boldsymbol{L}^{111}$ interact with each other through the superexchange processes, and the spatial configuration is controlled by the ratio $|\toffdiag/\tdiag|$, i.e., $\gamma$, as described above.
For $\gamma>\gamma_\mathrm{c}$, the ferro orbital ordering goes along with the 4in/4out-like spin structure under SOI, 
while the antiferro orbital ordering appears with the 2in--2out-like spin structure for $\gamma<\gamma_\mathrm{c}$. 
Our result clearly shows that the orbital-dependent transfers control the magnetism between the AFM and FM states together with the orbital ordering.

Figure~\ref{fig:hubbard}(b) shows the energy spectrum of the excited states.
For $\gamma>\gamma_\mathrm{c}$, the 4in/4out-like ground state is singled out and largely separated from other excited states because there is no nontrivial degeneracy in the ground state.
However, the gap vanishes toward $\gamma=\gamma_\mathrm{c}$, and then,
there appears a large number of low-energy states near the critical point due to the competition of the AFM and FM states.
Although the MLWF estimate $\gamma=0$ is in the 4in/4out-like region, it is close to the phase boundary, being consistent with the magnetic competition in Fig.~\ref{fig:pd}(a). 
This criticality tuned by the spin-orbital frustration gives the microscopic mechanism for the magnetic competition in Mo pyrochlores.
The competition may be robust for larger systems because of its local origin.

\subsection{Robustness of spin-orbital frustration}
\begin{figure}[h]
 \centering
 \includegraphics[width=.4\textwidth,clip]{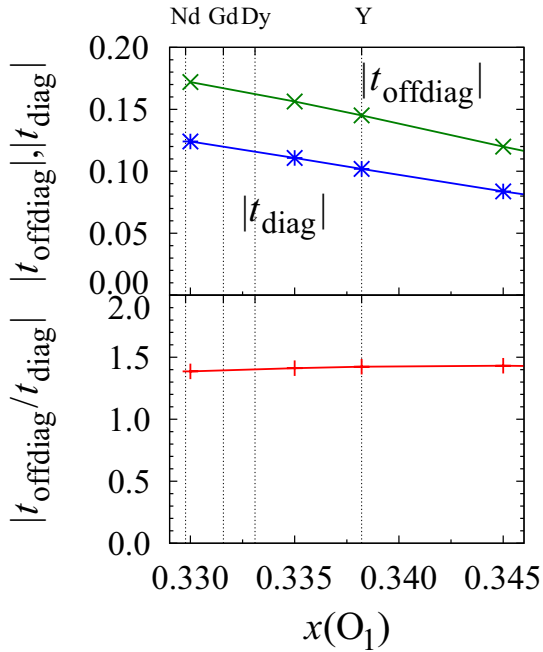}
 \caption{(color online). 
 The $u$ parameter $x(\mathrm{O}_1)$ dependences of the orbital diagonal/off-diagonal transfers $|\tdiag|$ and $|\toffdiag|$.
 Despite substantial changes in $|\tdiag|$ and $|\toffdiag|$, the ratio $|\toffdiag/\tdiag|$ is almost constant.
 The experimental values of $x(\mathrm{O}_1)$ for $A=$Y, Dy, Gd, and Nd~\cite{Reimers1988390,Moritomo01} are shown.
 }
 \label{fig:u-param}
\end{figure}

In real materials, the transfer integrals depend on the Mo-Mo distance and the angle of the Mo-O-Mo bond.
In particular, it is anticipated that the ratio between $\tdiag$ and $\toffdiag$ may depend on the latter. 
Here, we examine the $u$ parameter dependences of $|\tdiag|$ and $|\toffdiag|$ 
while fixing the lattice constant for Y$_2$Mo$_2$O$_7$. 
As shown in Fig.~\ref{fig:u-param}, although $|\tdiag|$ and $|\toffdiag|$ substantially change as $x(\mathrm{O}_1)$, $|\toffdiag/\tdiag|$ is almost constant at $\sim1.4$. 
In $A_2$Mo$_2$O$_7$, $x(\mathrm{O}_1)$ varies from $\sim 0.33$ to $\sim 0.34$ depending on $A$~\cite{Reimers1988390,Moritomo01}.
This suggests that the spin-orbital frustration discussed above will be commonly seen in the series of $A_2$Mo$_2$O$_7$. 
This might explain the robust SG behavior in the insulating compounds.

\section{Summary and discussion}\label{sec:summary}
In summary, we have investigated the spin and orbital states in the insulating phase of Mo pyrochlores.
Performing the electronic structure calculations by the fully relativistic LSDA+$U$ method, we have found the unconventional magnetic competition between AFM and FM states in the insulating region.
Through the analysis by using the generalized spin model, we have shown that the energy competition is explained by highly anisotropic magnetic interactions, being far distinct from the simple isotropic Heisenberg model.
We have revealed that the system is in the competing region between the AFM and FM phases.
Competing AFM and FM interactions lead to characteristic spin fluctuations in the finite-$T$ PM phase.
By the analyses of the three-orbital Hubbard model, we have revealed that the magnetic competition is tightly related to orbital ordering in the $4d^2$ configuration. Our results suggest that the spin and orbital frustration plays an important role in the insulating state in the Mo pyrochlores.

Our results provide a new insight into the puzzling SG behavior.
Orbital moments are polarized almost along the local [111] axes in the trigonal crystal field, being coupled with spin moments via SOI.
Under the severe competition between AFM and FM interactions, 
spin and orbital might freeze into a spin-orbital glass state at low $T$ in the presence of inevitable randomness in real materials.
This is in clear contrast to the conventional picture of SG where AFM NN exchange interactions are dominating~\cite{Saunders07,Andreanov10,Shinaoka10b,Shinaoka-LT2011}.
The renewed picture appears to be consistent with the diffuse scattering observed at FM points such as [000] and [222] in the recent neutron experiments for Y$_2$Mo$_2$O$_7$~\cite{Silverstein13}.

On the other hand, recently, the importance of magnetoelastic coupling to local lattice distortions in the SG behavior was pointed out experimentally~\cite{Booth00,Keren01,Sagi05,Greedan09,Ofer10} and theoretically~\cite{Shinaoka10b,Shinaoka-LT2011}.
Our results urge the reconsideration of SG behavior in the insulating $A_2$Mo$_2$O$_7$ by explicitly taking account of lattice distortions as well as orbitals.

%%%-----------------------------------------------------------------
\begin{acknowledgments}
We thank R. Kadono and H. Ohnishi for fruitful discussion.
We also thank T. Kosugi for the use of his computational code for fixing directions of spin moments in electronic structure calculations.
Numerical calculation was partly carried out at the Supercomputer Center, ISSP, Univ. of Tokyo. 
This work was supported by Grant-in-Aid for Scientific Research (No. 21340090, No. 22104010, No. 22540372, and No. 24340076), the Strategic Programs for Innovative Research (SPIRE), MEXT, and the Computational Materials Science Initiative (CMSI), Japan.
\end{acknowledgments}
%%%-----------------------------------------------------------------

%merlin.mbs apsrev4-1.bst 2010-07-25 4.21a (PWD, AO, DPC) hacked
%Control: key (0)
%Control: author (8) initials jnrlst
%Control: editor formatted (1) identically to author
%Control: production of article title (-1) disabled
%Control: page (0) single
%Control: year (1) truncated
%Control: production of eprint (0) enabled

\bibliography{ref}
\end{document}